# Assessing the Impact of Patent Attributes on the Value of Discrete and Complex Innovations

By

Mohd Shadab Danish[1], Pritam Ranjan[2], and Ruchi Sharma[1]

[1]Indian Institute of Technology, Indore

[2]Indian Institute of Management, Indore

Abstract

This study assesses the degree to which the social value of patents can be connected to the private value of patents across discrete and complex innovation. The underlying theory suggests that the social value of cumulative patents is less related to the private value of patents. We use the patents applied between 1995 to 2002 and granted on or before December 2018 from the Indian Patent Office (IPO). Here the patent renewal information is utilized as a proxy for the private value of the patent. We have used a variety of logit regression model for the impact assessment analysis. The results reveal that the technology classification (i.e., discrete versus complex innovations) plays an important role in patent value assessment, and some technologies are significantly different than the others even within the two broader classifications. Moreover, the non-resident patents in India are more likely to have a higher value than the resident patents. According to the conclusions of this study, only a few technologies from the discrete and complex innovation categories have some private value. There is no evidence that patent social value indicators are less useful in complicated technical classes than in discrete ones.

Contact: MOHD SHADAB DANISH*, PRITAM RANJAN†, and RUCHI SHARMA‡, *Assistant Professor in Economics, Dr BR Ambedkar School of Economics University, Bengaluru, India (Email: shadab@base.ac.in), †Professor in OM&QT, Indian Institute of Management Indore, India (Email: pritamr@iimidr.ac.in), ‡Professor in Economics, Indian Institute of Technology Indore, India (Email: ruchi@iiti.ac.in)

## 1. Introduction

The term "patent value" is sometimes misleading. As a result, many academicians sought to comprehend the elements of a patent and their determinants to answer the question of what distinguishes a valuable patent from a worthless patent (Moore, 2005; Bessen, 2008). Determining the value of a patent reveals the information about the necessity of the patent system. The work of Scherer et al., (1959), Mansfield (1986), and Levin et al., (1987) suggest that patent protection is required only in few sectors, most notably pharmaceuticals. A detailed examination of patent value (private value of patent) is necessary not just for tracking overall innovation but also for finding innovative companies and communicating the value of patents to investors like venture capital firms. Second, to understand the policy requirement to cull out low-quality patents from the economy to reduce economic in efficiency. Patent renewal data has long been used to calculate patent value estimations. Renewal costs are regarded to be a continuous investment burden for patent holders, making it expensive to retain a patent in force until it reaches its statutory life limit. Renewal costs must be paid on a regular basis, or the patent will expire before maturity. As a result, the value of maintaining patent protection over time is tied to the economic significance of the invention, and firms' renewal strategies can be viewed as a type of revelation mechanism.

The need for patenting as well as the approach for keeping a patent alive varies by technology and industry. As a result, it is crucial to comprehend the patent renewal decision, which is regarded as a proxy for patent value, as well as the behaviour of patent samples of complex and discrete technology. As discussed, there are predominantly two categories of innovation: cumulative technology innovation and discrete technology innovation. Discrete technology (also referred to as the single product innovation) refers to a single patent securing distinct products that can be brought to the market independently (Cohen, 2000). Patents are supposed to function as exclusivity rights in discrete technologies that generate revenue through in-house production or licensee royalty payments. For instance, in the pharmaceutical sector, the primary purpose of a patent is to prevent others from replicating the idea. Since simple alternatives for pharmaceutically active compounds are nearly never available, a patent fence (complementary patent) is not required for innovation to function exclusively.

Whereas the complex technology is described as a bundle of complementary patents which pulls different significance for inventors (Baron and Delcamp, 2012). Complementary patents are patents that, in most situations, function as "complimentary assets" (Teece, 1986). In order to bring a successful product to market, a licence for these complementary patents owned by various owners in the market is required. Unlike discrete invention patents, complementary patents are used as a bargaining chip. Generally, complementary patents do not award a monopoly over a marketable invention; rather, they grant blocking rights over a jointly controlled technology (Bessen, 2004). This could explain why economists in the economic literature have noted different patenting strategies in technology fields such as ICT and software compared to other technologies (Cohen et al. 2000). The ability of the patent system to fairly reward innovators for their contributions in discrete and complex technologies is a major policy concern. There must be some link between the private and societal value of patents for the patent system to generate socially optimal innovation incentives.

Several empirical studies have investigated the capacity of the social value of patents to predict private value (Lanjouw and Schankerman, 1999; Hall et al., 2005; Bessen, 2006). Nonetheless, these analyses have failed to uncover differences in the patent value of disaggregated discrete and complex technologies while maintaining the social value of patent measures constant. The link between the private and social value of patents determines the patent system's ability to compensate innovators for socially useful innovations. The patent system is at risk of encouraging strategic patenting on incremental contributions rather than innovative efforts and big innovations if the relationship is weakened.

This study investigates three aspects of patent valuation. First, we observe the relationship between measures of private and social value indicators for the aggregate sample. We observe the private value of patents by predicting the likelihood of renewal after every 0-6, 7-10, 11-15, and 16-20 years of patent terms. Our analysis includes three most commonly used social value indicators. (a), the *inventions' complexity* is measured by patent technology scope (4-digit IPC class), inventor size, and the grant lag. (b) the *filing strategy* includes the structure and quality of the drafted document (number of claims) and protecting the same patent in a different jurisdiction (family size). (C), the *ownership group* that is patent owned by Indians (resident) or foreigners (non-resident). The indicators effectively capture the various phenomena associated with this social value. For example, the number of claims may represent the patent's scope, and the number of inventors may signify the organization's collective effort. The ease of the patent office examination process and the international applicability of the invention are also conveyed by grant lag and family size respectively.

Second, we divide the sample into two categories: discrete and complex innovation and analyze whether social value indicators behave differently when it comes to explaining patents' private value. In this study, the technology is disaggregated into 33 categories (as per Schmoch 2008). In India, the renewal fee changes every 7th, 11th, and 15th year during the patent life. We argue that the change in the fee scale is likely to influence the patentee's renewal decision. Thus, the observable patent value is coded as 1 if a patent does not survive the first renewal fee hike (at the $7^{th}$ year), 2 if it expires between $7^{th}$ and $11^{th}$ year, 3 if the life of a patent is between $11^{th}$ and $15^{th}$ year, and 4 if it survives more than 15 years. We treat this coded patent value as our response variable in this study. Since the outcome variable is ordinal, we follow William (2006) and use the generalized ordered logit model to identify the valuable technologies for disaggregated complex and discrete technology fields. The findings of this study reveal a clear link between the social and private value of patents. Not all technologies in the discrete and complex technology categories are considered high value. This means that in India, just a few technologies are more valuable. This is one of main differences in quality between international and domestic patents.

The rest of the paper is organized as follows: Section 2 presents an overview of patent valuation literature and formulates working hypotheses. Data collection and the statistical models are discussed in Section 3. Section 4 presents the models, Section 5 summarizes the results by technologies and ownership category, and Section 6 concludes the chapter by discussing the implications of these findings for evaluating the need for IPRs in India.

## 2. Analytical Framework

The purpose of this section is to give a broad review of the literature and to describe our major contributions to the current state of the art. We start by describing the economic literature on measuring the private value of patents, with a focus on the use of patent value indicators. The focus is on the distinction between discrete and complex (cumulative) innovation. We also show how the current research adds to and expands on previous findings.

### 2.1. Private value of patents: patent value indicators

Economic research clearly distinguishes the private and social value of patents (Arrow, 1962; Trajtenberg 1990b). The social value of a patent is a bigger concept that includes the aggregate value created by the patent for social welfare. The concept of private value takes into account only the value-added of the patent for its owner: it can thus be defined as the depreciated sum of expected cash flows or the contribution of the patent to the market value of the owning firm. When we deduct the negative and positive externalities from the social value of a patent, we reach the private value of the patent (Bloom et al. 2013).

In a comprehensive review of the literature, Van Zeebroeck and Van Pottelsberghe de la Potterie (2011) emphasize the conceptual distinction between patent value determinants and indicators. The research setting determines whether a given variable is used as an indicator of patents' private or social value. Lanjouw and Schankerman (2004) and Bessen (2008) focus on patent renewal and litigation decision of patentee as a measure of the private value of a patent, as this decision involves an economic cost to its owner. The private value of a patent can be inferred indirectly from the decision-making of patentees (innovators/owners), which exposes the value they assign to their inventions. The remaining visible characteristics of patents, such as technological significance as measured by citations, patent breadth measured by claim number, and application approaches quantified by family size, are employed as causal factors in patent renewal and litigation decisions. These factors can be examined as causal determinants of private value if there is a correlation between social value indicators of patent and private value (measured in terms of renewal decision).

Observable patent attributes have often been used as a measure of patent social value in various studies (Baron and Delcamp, 2012). Hall et al. (2005) and Nagaoka (2005) investigate the relationship between patent indicators (patent quality indicators) that indicate patent social value and patentee private value. Further, the impact of patent "quality" indicators—determinants of patent value—on the likelihood of a patent being renewed is examined by Lanjouw and Schankerman (1999) and Thomas (1999). There is a continuous relationship between private and social value in patent valuation research, but there are also significant variances between technological domains (Baron and Delcamp, 2012).

The patent data's legal status gives essential information about the legal events, including expiration of a patent, renewal information, claims, change of legal identity, and other related information. Patent value indices constructed based on legal status are grant index (Zeebroeck, 2011), litigation index (Lanjouw and Schankerman, 2001; Hsieh, 2013), inventor index (Caviggioli et al., 2013), claim index (Trappey et al., 2012) and renewal index (Hikkerova et al., 2014). Since all these indices are based on literature, their applicability and validity can be verified. Hikkerova et al. (2014) study the patent life cycle in the European context argue that patent renewal information is critical because it reveals the benefit appropriated from the patent. Lanjouw and Schankerman (2001) use a litigation index to find that the cost of participating in litigation over IP assets reduces their worth as an incentive to invest in research. They also indicate that the likelihood of a patent being sued varies significantly between patents.

**Table 1 Patent value indicators**

| | | |
|---|---|---|
| Forward citations | The total number of citations received by subsequent patents | Indicates how important the patent is for future research. |
| Backward citations | The total number of referrals to earlier patents | Indicates how much of the existing prior art is used in the patent. |
| Number of claims | The patent's total number of priority claims. | Indicates the scope of the patent holder's claimed technology. |
| Family size | The number of worldwide patents for the same priority patent that have been filed. | Indicates that a patent is significant on a global scale and that the patent's validity has been confirmed by many patent offices. |
| Generality | Patent citations are dispersed among technological categories. | Indicates that the patent is based on multiple sources, implying that it is a fundamental rather than incremental breakthrough. |
| Originality | Dispersion of citing patents over technology classes | Indicates that the patent is significant for future research in a broad subject. |

Note: We compiled a list of the most commonly used social value indicators based on the literature The patent's value is connected to the particular attributes of technology and the R&D process, the nature of the market, and the competition pattern. The patent's role is higher when imitation is accessible, i.e., when the ratio between imitation costs and innovation costs is lower (e.g., chemicals, pharmaceuticals, machinery). Additionally, patents generally are more significant in the technologies where R&D is exceptionally capital concentrated and highly uncertain (pharmaceutical). When technical change is quick and the effective life of innovation is short, patents may not adequately reward innovators (semiconductors and software are good examples).

Additionally, patenting strategies range from discrete to cumulative invention, which could affect various indicators. for example, innovators are not interested in patenting their cumulative innovation in every single patent office in order to exclude potential imitation. Therefore, the patent family is higher in discrete technology patents compared to cumulative (complex) technology patents. Furthermore, in cumulative innovation, the presence of overlapping patents may create incentives to increase the number of claims, as a higher number of claims improves the likelihood that the patent would be relevant to future advances of a jointly held invention (Berger et al. 2012). The examined literature offers numerous reasons why patent indicators behave differently in discrete and cumulative innovation when estimating the private value of patents. Before studying discrepancies in the link between the private value of patents, it is necessary to test the consistency of patent indicators across different technological sectors.

**2.2. Private value of patent: Cumulative/complex versus discrete innovation**

The distinction between complex and discrete technologies is crucial to our research. Levin et al. (1987) were the first to make the distinction between complex and discrete technology, and it has since been the subject of a large body of research (Merges and Nelson,1990; Cohen, Nelson, and Walsh, 2000; Harhoff and von Graevenitz, 2009). The research indicates that a "complex" innovation field comprises of multiple complementary patents, often held by various inventors within one product. For example, a BluRay player incorporates several thousand patents held by various major players of the business. Contrary to that, only a few patents complete the product in discrete technologies that can be brought to the market independently. Generally, the IP for one product is held by one single owner.

The patent indicators vary with the complexity of the technology. For example, a patent web's density in a complex industry (cumulative innovation) generally affects the average number of claims and citations. Blind and Thumm (2010) find that patents identified as essential to the technological standard have more claims. The presence of overlapping patents could give incentives to raise the number of claims, as expanding the number of claims builds the odds of the patent being relevant to future innovation in similar technological areas (Baron and Delcamp, 2012). Important divergences between complex and discrete technologies have been revealed in a couple of empirical analyses of indicator performance. Hall et al. (2005) find that when innovation is cumulative, the quality of patent, in general, is less likely to correlate with its value. In a different approach, Lanjouw and Schankerman (2004) find that patent quality is the only underlying factor that could jointly affected by the number of claims, forward and backward citation, and size of the families. The value indicator could be anything that captures the patent's return from the inventor's perspective, such as renewal life and litigation.

Patenting data varies by nationality, so does the quality of innovation. Despite India's improving innovation potential (India's rank changed from 86 in 2013 to 46 in 2021), we have no convincing evidence that patents' social value indicators have any relation with the private value of the patent. This study considers patent's renewal years as a proxy for effective patenting (private value of patent). We attempted to determine whether patent value indicators such as social value, as defined by Baron and Delcamp (2012), have any impact on the renewal decision for complex and discrete technology patents. Furthermore, previous research has not employed the disaggregated technical classification to get micro variations across the technology sectors.

**2.3. Hypothesis development**

The patenting strategies are different from discrete to cumulative innovation, and therefore the social and private value of a patent can have a different link. In the cumulative invention, for example, not all complementary components of technology should be patented in every office to prevent future imitation. As a result, discrete patent families are larger than cumulative patent families. Furthermore, overlapping patents in cumulative innovation contain more claims, as an increased number of claims enhances the likelihood that the patent may be relevant to future developments of a jointly held invention (Berger et al. 2012). As a result, the cumulativeness of innovation in a single technical field is perceived to move indicators upwards or downwards, whereas variance within a sample is unaffected (Baron and Delcamp, 2012). This has been well proven in developed country innovation; but, without empirical evidence, we cannot verify the same for developing countries. We proposed three hypotheses to better understand the relationship between patent social and private value across discrete and complex technology (cumulative innovation).

To begin with, we first conduct a test of hypothesis to verify whether the categorization of technologies in discrete versus complex reveals any value differences. This hypothesis will provide light on India's patent general quality phenomenon. It will assist policymakers in more effectively promoting a strategic patenting (cumulative innovation) or discrete innovation policy.

> *$H1_0$: There is no difference in the value of discrete and complex innovation*
>
> *$H1_a$: Value of discrete and complex innovation differs*

The second S discusses the importance of social value indicators in explaining the private value of patents. The findings of patent value in industrialized countries may suggest that, while the indicators still reveal a similar "quality" component in complex technology classes, they provide less consistent results in areas where innovation is more cumulative. We want to see if the general conclusion holds true in India.

> *$H2_0$: Social value of patent explain complex and discrete technology patent uniformly*
>
> *$H2_a$: Social value of patent explain complex and discrete technology patent differently*

The study's third objective is to investigate the differences in value between various types of ownership. This will highlight the country's capacity for innovative thinking. The knowledge of patent quality by ownership (resident and non-resident) will provide evidence for future policy adjustments to improve the country's quality, if necessary.

> *$H3_0$: The value of patents is uniform across the resident and non-resident patents of discrete and complex technology*
>
> *$H3_a$: The value of patents is not uniform across the resident and non-resident patents of discrete and complex technology*

The hypothesis test results are discussed in Section 5.

## 3. Data Description and Variable Selection

This section focuses on the sources of data, processing, and model specification.

### 3.1. Data

We collected patent-wise information from Indian Patent Office (IPO), New Delhi, India, for all granted patents filed/applied between 1st January 1995 and 31st December 2002. Using the IPO patent search engine (http://ipindiaservices.gov.in/publicsearch), we retrieved patent application numbers for each year. On the IPO website, each patent number is searched separately to obtain patent-level information such as the number of claims, inventor size, renewal years, and so on. The total number of patents applied at IPO by resident and non-residents during the sampling period was 69,658, out of which only 26,362 patents were granted. Furthermore, only 21,562 patents contained complete information on the renewal time and patent social value indicators used in this study.

All patents are partitioned into subsets corresponding to technology areas using the Schmoch's (2008) classification (as updated in 2010 and 2011), which relies on the International Patent Classification (IPC) codes contained in the patent documents. The five major sectors: electrical, instruments, chemistry, mechanical, and "otherfield" are further subdivided into 33 sub-technology groups[1]. To avoid double-counting of patents, this study uses the first classification codes of each patent to determine the technology class.

---

[1] The detailed list of the IPC classes contained in each technology field can be found at www.wipo.int/ipstats/en/statistics/patents/pdf/wipo_ipc_technology.pdf

We bisect these 33 technology areas according to the definition of complex and discrete technologies suggested by Cohen et al. (2000) and Graevenitz et al. (2011) to assign 1 if it falls in the complex category and 0 if discrete (see Table 2).

**Table 2: Classification of technologies into discrete and complex innovations**

| Classification | Technology Area |
|---|---|
| Discrete (12) | Organic fine chemistry (2034), Biotechnology (532), Pharmaceuticals (532), Macromolecular chemistry polymers (707), Food chemistry (318), Basic materials chemistry (1295), Materials metallurgy (1133), Surface technology and coating (376), Chemical engineering (1158), Handling (593), Textile and paper machines (828), Furniture and games (158) |
| Complex (21) | Electrical machinery apparatus and energy (1225), Audio-visual technology (702), Telecommunications (960), Digital communication (306), Basic communication processes (220), Computer technology (818), Semiconductors (149), Optics (213), Measurement (575), Analysis of biological materials (119), Control (187), Medical technology (775), Environmental technology (263), Machine tools (634), Engines pumps turbines (701), Other special machines (664), Thermal processes and apparatus (357), Mechanical elements (571), Transport (705), Other consumer goods (351), Civil engineering (382) |

Note: Data collected from Indian Patent Office (IPO)

Out of the total sample of 21,562 patents, 49.55% (10,685) patents belong to the complex technology fields, and 50.44% (10,877) of the patents fall under discrete technology categories. As per our results, since there is a significant difference between the discrete and complex technologies with respect to the patent value estimation, we fitted models for both discrete and complex technologies separately. According to the ownership dummy variable, the share of non-resident patents is around 83 percent (18078) (and the number of resident patents is 3484). Next, we discuss patent social value indicators (measures) for which the data were collected from the IPO website.

### 3.2. Patent Social Value Indicators

In order to determine whether there are any observable indicia of a patent value or lack of value, we estimate the likelihood of renewal across a large number of variables. In particular, we examine the role of the following characteristics which may influence the likelihood of a patent owner failing to pay the maintenance fees: (a) number of claims, (b) family size, (c) technology scope, (d) grant lag, (e) the number of inventors listed in the patent (inventor size), and (f) ownership dummy for a foreign or Indian resident patentee. We now briefly describe each of these variables:

(a) Claims: A patent has a bunch of claims that portray what is ensured by the patent. The principal claim explains the fundamental novel highlights of the innovation in their broadest structure, and the subordinate claims describe a feature of the innovation. In this article, we take the total number of claims to determine the renewal decision factor. The patentee intends to increase the claims as much as possible to get a maximum incentive from the innovation.

(b) Family size: A group of patents protecting the same innovation constitutes a 'family' (also called parallel patents). Filing and maintaining a patent in different countries is associated with high costs, and only a fraction of patents seek protection outside their home market. Therefore, the family size (the number of jurisdictions (patent offices) in which a patent is filed) indicates the importance of a patent.

(c) Technology scope: The examiner assigns each patent a 9-digit code based on the IPC classification system. We use the 4-digit subclass count in a patent to describe the technology scope—the broader the technology, the higher the count of the 4-digit subclass of a patent.

(d) Grant-lag: The grant lag is defined as the time elapsed between the filing and grant date of a patent. Harhoff and Wagner (2009) and Régibeau and Rockett (2010) find evidence of an inverse relationship between patent value and the grant lag. We investigate the impact of grant-lag on patent value in the Indian context.

(e) Inventor size: We use the inventor count given in the patent data to indicate the project's size and complexity (Gambardella et al., 2006).

(f) Ownership: A patent filed at IPO and assigned to India is called a resident patent (coded as 0), but if it is assigned to another country, then it is called a non-resident patent (coded as 1). Table 3 outlines the usage of these patent social value indicators in the existing literature. Descriptive statistics on the patent social value indicators are presented in Section 5.1.

**Table 3: Summary of the response variable (renewal level) and independent variables (patent social value indicators) used in the regression models.**

| Variable | Description | References |
|---|---|---|
| **Renewal level (RL)** | Each patent is classified in one of the four categories (1, 2, 3, and 4) based on the number of years a patent has been renewed (see Table 1). | Reitzig (2004); Moore (2005); Bessen (2008) |
| **Family Size (FS)** | The number of jurisdictions a patent is filed in. | Kabore and Park (2019); Harhoff et al. (2003) |
| **Number of Claims (NC)** | Number of innovations claimed in a patent. | Reitzig (2004); Caviggioli et al. (2013) |
| **Grant Lag (GL)** | Time elapsed between filing and grant date. | Harhoff and Wagner, (2009) |
| **Technology Scope (TS)** | Number of technological domains a patent belongs to. Four-digit IPC-code captures the information. | Squicciarini et al. (2013); Lerner (1994) |
| **Inventor Size (NI)** | The number of inventors involved in a patent. It also measures the R&D size and scale of a patent. | Kiehne and Krill (2017) |

Note: The social value indicators used in this study were chosen based on the literature.

## 4. Empirical Models

The response variable in our models is defined by the four-level ordered categorical variable that characterize the patent renewal life guided by India's renewal fee structure (referred to as "renewal level" in Table 1). Given that the dependent variable is divided into more than two categories with a meaningful sequential order, the most intuitive and popular choice of the model is an ordinal logit regression model which efficiently analyses the patent valuation with respect to different patent social value indicators and technological domains.

### 4.1. Proportional odds model

A common approach for modelling such an ordinal response is to use the proportional odds model (POM) developed by McCullagh (1980), also known as the cumulative logit regression model. If the response variable $Y$ (here, the renewal level) has $J$ ordered categories ($J = 4$, as per Table 1), then the model is given by (Long and Cheng 2004)

$$log\left(\frac{Pr\ (Y \leq j\ |\ x)}{Pr\ (Y > j\ |\ x)}\right) = \tau_j - x'\beta, j = 1,2,\dots,J-1, (1)$$

where $j$ represents the renewal level (i.e., $j = 1,2,3$), $\beta$ is the vector of regression coefficients corresponding to the input vector (i.e., the patent social value indicators), and $\tau_j$ is the cutoff effect between response category boundaries. The negative and positive signs of $\beta$ coefficients are interpreted similarly as in the OLS and binomial logistic regression. The proportional odds model assumes regression coefficient $\beta$ to be the same across the three logit equations.

On several occasions, the proportionality assumption is violated, and thus, the results obtained are biased. One of the most popular method to test the proportionality assumption is proposed by Brant (1990), which uses an omnibus chi-square test. A significant test statistic would indicate that the parallel regression assumption has been violated, which happens to be the case here for a few patent social value indicators. Results are presented in Section 5.2. Consequently, we adopted an alternative model, the Generalized ordered logit model (GOLM), suggested by Williams (2006; 2016).

**4.2. The generalized ordered logit model**

The main idea here is that both the intercept and the regression coefficient vector $\beta$ (corresponding to the patent social value indicators) can vary across the $J$ categories of response (i.e., renewal level). The model statement is given by

$$log\left(\frac{Pr\ (Y \leq j\ |\ x)}{Pr\ (Y > j\ |\ x)}\right) = \alpha_j - x_j'\beta_j, j = 1,2,\dots,J-1, (2)$$

where $J$ is the number of outcome categories of the ordinal dependent variable, $\alpha_j$ if the relative cutoff effect for category $j$ and $\beta_j = (\beta_{j1}, \beta_{j2}, \dots \beta_{jk})$ correspond to the regression coefficients with respect to the $k$ independent variables (patent social value indicators and technological indicators). Note that the proportional odds model (POM) is a special case of GOLM, where the regression parameter vector $\beta_j$ are the same for each categorical level $j = 1, \dots, J-1$. The econometric model applied in this study simplifies the real-world process and contains the salient feature of patent valuation phenomena.

This exploratory study examines the correlations between an individual patent-protected invention value and patent social value indicators. The following is a specification that provides for a preliminary empirical test of the hypotheses outlined above:

Hypothesis 1:

$Privatevalue_i = \beta.socialvalue_i + \delta.complextech_i + \gamma.ownership_i + \varepsilon_i$ \qquad (3)

Hypotheses 2 and 3:

$Privatevalue_i = \alpha.Disaggregatedtech_i + \beta.socialvalue_i + \gamma.ownership_i + \varepsilon_i$

where, "Privatevalue" is coded in 1, 2, 3, and 4, *i* denotes patents, "Disaggregatedtech" category includes dummy for all 33 technologies, the "socialvalue" of the patent denotes all the patent attributes such as a number of claims, inventor size, etc., "ownership" dummy explains if the patent is resident or non-resident, and "complextech" dummy explains if patent belongs to discrete technology category or complex technology category.

## 5. Empirical Results

We start by summarizing the data from various standpoints and then discuss the two logit models (POM and GOLM). We particularly focus on assessing technological domains in influencing the patent value measured via the "renewal level". The hypotheses listed in Section 2.2 are also tested and discussed here.

### 5.1. Descriptive statistics

The most basic summary (mean and standard deviations) of the patent social value indicators for the samples in discrete and complex technologies are presented in Tables 4 and 5, respectively. Recall that the total patent data of 21,562 consists of 10,685 discrete technology and 10,877 complex technology patents.

**Table 4: Summary statistics and correlation matrix of patent social value indicators for discrete technology patents**

|  | Claims | Inventor size | Family size | Technology Scope | Grant lag |
|---|---|---|---|---|---|
| Claims | 1 | | | | |
| Inventor size | 0.0444* | 1 | | | |
| Family size | 0.1781* | 0.0877* | 1 | | |
| Technology Scope | 0.2209* | 0.1584* | 0.6147* | 1 | |
| Grant lag | -0.0231 | 0.0573* | -0.0822* | -0.0858* | 1 |
| Mean | 12.68 | 2.93 | 18.82 | 8.17 | 8.05 |
| Median | 10 | 2 | 15 | 4 | 8 |
| Min | 1 | 1 | 1 | 1 | 1 |
| Max | 366 | 22 | 315 | 183 | 20 |
| Std Deviation | 12.21 | 2.06 | 20.48 | 11.35 | 2.58 |
| Observations | 10,685 | 10,685 | 10,685 | 10,685 | 10,685 |

Note: Estimated two-tail tests at 5% level of significance (i.e., 95% confidence level).

**Table 5: Summary statistics and correlation matrix of patent social value indicators for complex technology patents**

|  | Claims | Inventor size | Family size | Technology Scope | Grant lag |
|---|---|---|---|---|---|
| Claims | 1 | | | | |
| Inventor size | 0.0814* | 1 | | | |
| Family size | 0.1503* | 0.0620* | 1 | | |
| Technology Scope | 0.1635* | 0.0925* | 0.7025* | 1 | |
| Grant lag | -0.0639* | 0.0512* | -0.0911* | -0.0696* | 1 |
| Mean | 13.63 | 2.32 | 15.81 | 5.57 | 8.31 |
| Median | 10 | 2 | 13 | 4 | 8 |
| Min | 1 | 1 | 1 | 1 | 1 |
| Max | 246 | 22 | 381 | 125 | 19 |
| Std Deviation | 13.62 | 1.72 | 19.18 | 6.13 | 2.44 |

| Observations | 10,877 | 10,877 | 10,877 | 10,877 | 10,877 |

Note: Estimated two-tail tests at 5% level of significance (i.e., 95% confidence level).

A few notable findings are as follows. The average grant lag for discrete technology patents filed during 1st January 1995 and 31st December 2002 at IPO is 8.05 years, whereas, in the complex technology category, the grant lag is slightly higher (8.33 years). In recent times, India's average grant lag has reduced to 64 months (5 years), which is still higher than 22 months in China and European patent offices and 24 months in the US (WIPO, 2019).

The patenting strategies are not quite the same for discrete and complex innovation. For example, in complex/cumulative innovation, not all complementary parts of innovation should be patented in every office to exclude possible imitation. Along these lines, the average family size is bigger in discrete (18.82) than in complex (15.81) innovations. Moreover, the presence of overlapping patents in cumulative innovation could give motivations to raise the number of claims, as expanding the number of claims builds the odds of the patent to be applicable to future improvements of a jointly held innovation (Berger et al., 2012). The descriptive statistics also validates Berger et al., (2012) argument on the average number of claims in the complex innovations (13.63) being higher than the discrete innovations (12.68). In contrast, patents filed at the Japanese Patent Office (JPO) and European Patent Office (EPO) have average claims of 10.4 and 14.7, respectively. In China, the average number of claims is 8.1 (IP5 Statistics Report, 2017). Broad claims suggest that the patent could successfully block the access to incremental innovation based on original technology, and thus, it is one of the important determinants of patent value.

The pairwise correlation matrices in Tables 4 and 5 show that no two patent social value indicators are highly linearly related. We also computed the VIF (variance inflation factor) values (see Table 6) which are very small (close to 1) and hence reject multicollinearity among the predictors.

**Table 6: Variance inflation factor (VIF) values of the patent social value indicators for all 21,562 patents**

| Variable | VIF | Tolerance (1/VIF) |
|---|---|---|
| Technology Scope | 1.69 | 0.59 |
| Family Size | 1.65 | 0.6 |
| Claims | 1.04 | 0.96 |
| Inventor Size | 1.03 | 0.97 |
| Grant Lag | 1.01 | 0.99 |

| | Mean VIF | 1.29 | - |

Note: 1 value of VIF present no correlation; between 1 to 5 moderately correlated, greater than 5 highly correlated.

We now look at the distribution of patents and the renewal behaviour. Table 7 presents the empirical distribution of patents according to the renewal level for discrete, complex, and combined categories. The table reveals a somewhat increasing trend, which is expected because if someone has filed a patent, then the patent is likely to be worthy enough to be renewed for at least a few years. Some studies have found that the expected renewal life of a patent is shorter in developing countries than developed countries (e.g., Gupeng and Xiangdong, 2012). This may be attributed to the fact that a bulk of innovations are of incremental value.

**Table 7: Distribution of patents in different response categories**

| Patent life | Renewal level | Discrete (%) | Complex (%) | Combined (%) |
|---|---|---|---|---|
| 0 to 6th year | 1 | 16.98 | 16.87 | 17.14 |
| 7th year to 10th year | 2 | 12.37 | 12.09 | 12.66 |
| 11th year to 15th year | 3 | 27.28 | 26.91 | 26.75 |
| 16th year to 20th year | 4 | 43.37 | 44.13 | 43.46 |
| | Total patents | 100 (10,685) | 100 (10,877) | 100 (21,562) |

Note: Renewal years information is obtained from IPO. Renewal level 1 to 4 shows the proportion of patent expired in each category.

Further analysis of patents with respect to the major technological fields reveals that electrical patents are more likely to be maintained by their owners. In contrast, mechanical patents expire more often at an early age (see Table 8). Moreover, a high percentage of patents belonging to instruments and "otherfield" have never been renewed by their owners'. Around 17.14% of the total patents across different technologies have never been renewed, and 56.5% of patents expire before the 16th year. This implies most of the learning from the patent happens in the early stage of the patent application. Thus, a significant number of patents expire without completing 20 years of their lifetime.

**Table 8: Patent survival rate in different technology fields**

| | Never Renewed | 3rd to 6th | 7th to 10th | 11th to 15th | 16th to 20th |
|---|---|---|---|---|---|
| Electrical | 15.96 | 0.39 | 12.44 | 23.45 | 47.76 |
| Instruments | 19.32 | 0.43 | 11.99 | 25.63 | 42.64 |
| Chemistry | 15.46 | 0.29 | 12.01 | 27.12 | 45.12 |
| Mechanical | 17.95 | 0.51 | 14.05 | 29.43 | 38.06 |
| Others | 22.25 | 0.45 | 14.04 | 26.18 | 37.19 |
| Total | 18.18 | 0.38 | 12.66 | 26.75 | 43.46 |

are included in each technological category. All survival rates are expressed as a percentage.

Figure 4 depicts the patent survival rate for different technology categories. It is clear from Figure 4 that the number of patents that expire between 0-2 years of patent life is highest in "otherfield" category and lowest in chemistry. As expected, the differences in patent survival rates decline across the technology as it approaches the 16th year of their life.

**Figure 4: Patent Survival Curve for Different Technology Group**

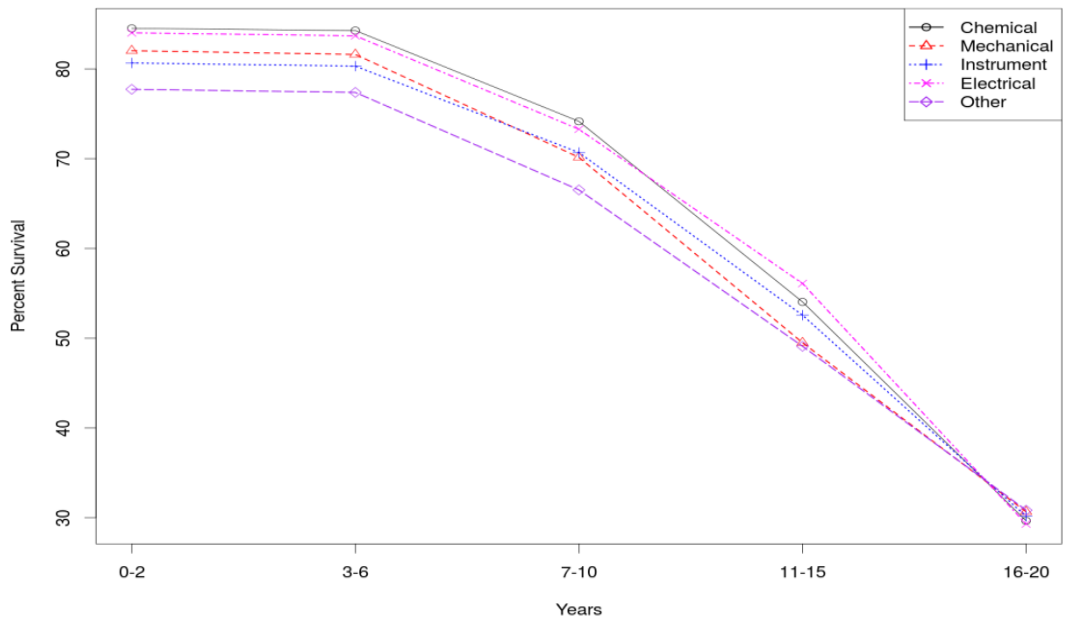

## 5.2. Ordered logit regression model results

Here, we fit the proportional odds model (POM) for the combined sample to quickly check hypothesis H1 (difference between the impact of discrete and complex technological innovations on patent values). We study the technology innovations separately for the two classes (see Table 9). We also conducted a Brant test (1990) for validating the parallel regression assumption in POM.

**Table 9: Regression coefficients of proportional odds model and Brant test results of discrete and complex technology**

|  | Discrete Technology | | Complex Technology | |
|---|---|---|---|---|
| Explanatory variables | Coefficient | Brant test | Coefficient | Brant test |
| Claims | -0.04** (0.03) | 2.57 | 0.03 (0.02) | 23.89*** |
| Inventor size | 0.37*** (0.04) | 5.63* | 0.29*** (0.04) | 6.74** |
| Family size | 0.18*** (0.03) | 11.87*** | 0.23*** (0.03) | 3.39 |
| Technology scope | -0.08** (0.03) | 2.41 | -0.05 (0.04) | 18.99*** |
| Grant lag | 0.85*** (0.06) | 220.91*** | 0.71*** (0.07) | 342.87*** |
| Ownership | 0.23*** (0.05) | 39.16*** | 0.09 (0.06) | 3.67 |
| Cut1 | 1.03 (0.16) | | 0.91 (0.18) | |
| Cut2 | 1.74 (0.16) | | 1.67 (0.18) | |
| Cut3 | 2.91 (0.16) | | 2.80 (0.18) | |
| Observations | 10,685 | | 10,877 | |

**Notes:** Here, ***, **, and * denote p < 0.01, p < 0.05 and p < 0.10, respectively.

A significant test statistic provides evidence that the parallel regression assumption has been violated (more specifically, inventor size, family size, and ownership dummy for discrete technology; and claims, inventor size, technology scope, and grant lag for complex technologies). The large values of chi-square test statistics (shown in Table 9) suggest that several explanatory variables violate the proportionality (or parallel) assumption. Thus, we investigate the alternative model: generalized ordered logit model (GOLM).

### 5.3. Generalized ordered logit regression model results

GOLM assumes that the regression coefficient vector $\beta_j$ may vary across different logit equations with respect to different "renewal levels" $j = 1,2,3$. The results are presented in three panels corresponding to $P(Y \leq 1), P(Y \leq 2)$, and $P(Y \leq 3)$. That is, the first panel analyzes the model for "renewal level" category 1 vs. 2, 3, and 4; the second panel presents the regression coefficients for "renewal level" category 1, 2 vs. 3, 4; and so on. For a quick check, we fitted this model for the complete data set using complex vs. discrete technology dummy and found consistent result (as in Table 9) that the two innovation categories are significantly different, and the complex innovations in India are less valuable as compared to discrete innovations. Subsequently, we analyze the two technologies separately. Tables 10 and 11 present the model results for complex and discrete technologies respectively.

**Table 10: Analysis of Generalized ordered logit regression model (GOLM) for complex innovation (Reference category: computer technology)**

|  | 1 vs 2 3 4 | 1 2 vs 3 4 | 1 2 3 vs 4 |
|---|---|---|---|
| Explanatory variables | Coefficient | Coefficient | Coefficient |
| Claims | -0.05 (0.04) | -0.02 (0.03) | 0.05* (0.03) |
| Inventor size | 0.38*** (0.06) | 0.32*** (0.05) | 0.23*** (0.05) |
| Family size | 0.24*** (0.04) | 0.29*** (0.04) | 0.23*** (0.03) |
| Technology scope | 0.06 (0.06) | 0.00 (0.05) | -0.16*** (0.04) |
| Grant lag | -0.30*** (0.10) | 1.23*** (0.09) | 0.76*** (0.08) |
| Ownership | 0.01 (0.08) | 0.01 (0.07) | 0.13* (0.07) |
| Electrical machinery, apparatus, energy | -0.17 (0.11) | 0.03 (0.10) | -0.13*** (0.09) |
| Audio-visual technology | 0.12 (0.14) | 0.21** (0.11) | 0.30 (0.10) |
| Telecommunications | 0.22* (0.12) | 0.38*** (0.11) | 0.53*** (0.10) |
| Digital communication | 0.11 (0.19) | 0.42*** (0.15) | 0.67*** (0.14) |
| Basic communication processes | -0.08 (0.20) | 0.48*** (0.18) | 0.39*** (0.15) |
| Semiconductors | -0.09 (0.24) | 0.36* (0.20) | 0.20 (0.18) |
| Optics | -0.14 (0.20) | 0.07 (0.17) | 0.10 (0.16) |
| Measurement | 0.15 (0.14) | 0.34*** (0.12) | 0.13 (0.11) |
| Analysis of biological materials | -0.53** (0.24) | -0.21 (0.20) | -0.21 (0.21) |
| Control | 0.21 (0.23) | 0.31* (0.18) | 0.38** (0.16) |
| Medical technology | -0.34*** (0.12) | -0.11 (0.11) | -0.15 (0.10) |
| Environmental technology | 0.22 (0.19) | 0.30*** (0.16) | -0.13 (0.15) |
| Machine tools | 0.06 (0.14) | 0.31*** (0.12) | 0.11 (0.11) |
| Engines, pumps, turbines | 0.17 (0.14) | 0.26** (0.11) | -0.12 (0.11) |
| Other special machines | -0.19 (0.13) | 0.07 (0.11) | -0.09 (0.11) |
| Thermal processes and apparatus | -0.12 (0.16) | 0.22 (0.14) | 0.12 (0.13) |
| Mechanical elements | -0.02 (0.14) | -0.05 (0.12) | -0.21* (0.11) |
| Transport | 0.15 (0.14) | 0.01 (0.11) | -0.44*** (0.11) |
| Other consumer goods | -0.13 (0.16) | -0.03 (0.14) | -0.16 (0.13) |
| Civil engineering | -0.27* (0.15) | -0.02 (0.13) | -0.11 (0.13) |
| _cons | 1.28*** (0.26) | -3.02*** (0.23) | -2.78*** (0.22) |

**Notes**: Here, ***, **, and * denote $p < 0.01$, $p < 0.05$ and $p < 0.10$, respectively. Numbers in parentheses are standard errors.

The second hypothesis of the paper is to find the impact of readable patent social value indicators on the social value indicator of the patent. Some of this study concerning the second hypothesis suggests a significant impact of patent level characteristics on the private value of the patent. A few observations from Tables 10 and 11 are as follows.

(i) The coefficients of inventor size are positive across the three panels (corresponding to the renewal level cutoff), suggesting that a patent with many inventors and large family size is more likely to be maintained to a full term in both discrete and complex innovations. Since patents are territorial in nature, the invention is only protected in countries where patentees see the potential benefits. The existence of a patent in one country has no meaning in the legal system of other countries. The theoretical argument has been well established in the literature (Basberg, 1987; Putnam, 1996; Lanjouw et al., 1998). Reitzig (2004) conducted interviews with patent attorneys that confirmed that a patent's value is associated with the family size.

**Table 11: Analysis of Generalized ordered logit regression model (GOLM) for discrete innovation (Reference category: pharmaceutical)**

|  | 1 vs 2 3 4 | 1 2 vs 3 4 | 1 2 3 vs 4 |
|---|---|---|---|
| Explanatory variables | Coefficient | Coefficient | Coefficient |
| Claims | -0.04* (0.03) | -0.04* (0.03) | -0.04* (0.03) |
| Inventor size | 0.33*** (0.04) | 0.33*** (0.04) | 0.33*** (0.04) |
| Family size | 0.20*** (0.03) | 0.20*** (0.03) | 0.20*** (0.03) |
| Technology scope | -0.04 (0.04) | -0.10*** (0.04) | -0.14*** (0.04) |
| Grant lag | -0.05 (0.09) | 1.31*** (0.08) | 0.97*** (0.07) |
| Ownership | 0.10 (0.07) | 0.13** (0.06) | 0.47*** (0.06) |
| Organic fine chemistry | 0.00 (0.08) | -0.05 (0.07) | -0.21*** (0.07) |
| Biotechnology | 0.19** (0.10) | 0.19** (0.10) | 0.19** (0.10) |
| Furniture, games | -0.65*** (0.15) | -0.65*** (0.15) | -0.65*** (0.15) |
| Macromolecular chemistry, polymers | -0.17** (0.08) | -0.17** (0.08) | -0.17** (0.08) |
| Food chemistry | -0.13 (0.15) | -0.05 (0.13) | 0.21* (0.13) |
| Basic materials chemistry | 0.03 (0.07) | 0.03 (0.07) | 0.03 (0.07) |
| Materials, metallurgy | 0.14* (0.08) | 0.14* (0.08) | 0.14* (0.08) |
| Surface technology, coating | -0.05 (0.11) | -0.05 (0.11) | -0.05 (0.11) |
| Chemical engineering | -0.02 (0.07) | -0.02 (0.07) | -0.02 (0.07) |
| Handling | -0.21** (0.09) | -0.21** (0.09) | -0.21** (0.09) |
| Textile and paper machines | -0.26*** (0.10) | -0.37*** (0.09) | -0.46*** (0.09) |
| _cons | 0.96*** (0.23) | -2.57*** (0.20) | -3.20*** (0.19) |

**Notes:** Here, \*\*\*, \*\*, and \* denote $p < 0.01$, $p < 0.05$ and $p < 0.10$, respectively. Numbers in parentheses are standard errors.

Our study in the Indian context finds the positive impact of family size on the patent's renewal life. This implies that patents in other countries of the same invention are effective for identifying valuable patents.

(ii) The impact of the number of claims on the patent value is negative and significant in the discrete technology category, whereas it is found positive in a complex technology category. The result obtained reveals different patenting strategies in the discrete and complex technology category. For instance, overlapping patents in the cumulative/complex innovation provide an incentive to raise the number of claims, as increasing the number of claims raises the chance of the patent to be relevant to future developments of a jointly held technology (Berger et al. 2012).

(iii) Technology scope coefficients in both discrete and complex technology samples have a negative sign. This implies that patents having broader technology scope are less likely to fall in the higher value category. It also suggests that patent breadth, an indicator of technology broadness in India, is not as important as it is found in the developed countries context (Putnam 1996). The patent with higher grant lag is more likely to fall in the high-value category in both samples.

(iv) We further test the third hypothesis that says any differences in the value of resident patents in both discrete and complex technology categories. Foreign-owned patents (non-resident) in discrete and complex technology categories have a higher value than domestic patents. The greater difference is that non-resident patents are more likely to have a higher value in the discrete category than complex patents. The overall model for ownership of the patent also finds similar results.

(v) Returning to the study's main focus, we now discuss the effect of technological domains on the value of patents measured via its renewal length. In the discrete technology group, we used pharmaceutical as the technology baseline for the regression (randomly chosen). Biotechnology, food chemistry, and material metallurgy patents are more valuable than the baseline category. On the other side, organic fine chemistry; macromolecular chemistry polymers; handling; textile, and paper machines are less valuable than the baseline category. We also find that many technology fields -furniture, games; basic materials chemistry, chemical engineering; and surface technology coating are insignificant and have a similar impact on patent value as compared to the baseline category.

Among the complex technology category, we selected computer technology as the base category. It shows that Audio-visual technology; telecommunications; digital communication; basic communication processes, semiconductors, measurement; Environmental technology; Machine tools; Engines, pumps, turbines; control; are more valuable compared to the baseline category. The greatest difference is that semiconductors, measurement; Environmental technology; Machine tools; Engines, pumps, turbines are more likely to fall in panel 2; and audio-visual technology; telecommunications; digital communication; basic communication processes; control; panel 3 (higher value category).

We also find that electrical machinery, apparatus, energy; analysis of biological materials; medical technology; mechanical elements; transport is less likely to fall in the higher value category. The greatest difference is that analysis of biological materials and medical technology patents is more likely to fall in the lower value category than the baseline category.

## 6. Conclusion

This study is devoted to assess the link between the private and social value of patents (patent value determinants) across technological classes and especially between complex and discrete technologies. The model presented in this paper improves on the usual approach to patent renewal decisions by allowing for the integration of additional patent social value indicators. We ensure that measures of private and social value are consistent across the samples to be compared in this study. The disparities in the value have been frequently linked to the implications of more or less cumulative innovation, as theoretical reasons indicate a weaker link between the private and social value of patents when innovation is cumulative. However, the outcome of this study permits some generalities. The social value of patents measured in terms of invention complexity and fling strategy is strongly linked to the private value of patents in India across discrete and complex innovation patents. Thus, the social value of a patent is a good indicator in predicting the private value of patents. One distinction is that in complex technology, a broader technological claim is beneficial to negotiation therefore inventor renews it for a longer period of time, whereas in discrete technology patents, higher claims actually detract from the strength and value of the patent. unlike Baron and Delcamp (2012) this study concludes that the social value of patents is highly linked with the cumulative and discrete technology patents in India.

The patent protection is unnecessary in some discrete and complex technology, yet it may be required in others. In India, the benefits of collaboration should be stressed because collaborative patents are more likely to be valuable. Furthermore, the Indian patent system should concentrate on reducing the inefficiencies caused by absurd patenting through a rigorous assessment process before awarding. Another significant conclusion in terms of discrete and complex technology is that in India, only a few discrete technology patents are more valuable than cumulative/complex inventions. This implies that process innovation continues to dominate the Indian market, which is possible because product innovation was not permitted before 2005[2] in the pharmaceutical sector. We relate our result with the patent portfolio theory which says that organizations accumulate an enormous number of related patents and use them as a bargaining chip in negotiations with other patent owners. Further, the patents are useful in India, and the increase in patenting in highly cumulative technological fields is not a cause for concern. Another conclusion of this study supports the preceding claim that patents are heterogeneous both within and between samples. As a result, only a few technologies contribute considerably, while a vast number of innovations are unimportant in the Indian setting. To capture the true impact of a study using a patent as an indicator, the quality element of the study must be reevaluated.

---

[2] Product patent regime started with amendment in Indian Patent Act, 1972.

There are a few limitations of this examination that need consideration in future work. This research takes into account precise patent-level data. However, it omits some key details from the patent, such as forward citation, backward citation, and Indian inventor proportionality. Another important piece of information on "market maturity" with recent data (after 2005) can help with valuation. In the future, samples could be divided into three categories based on assignee groups, such as patents owned by individuals, institutions, and companies. This may aid the reader in comprehending the value disparities among the assignee group. This will provide strong evidence for the relevance of the patent system in developing countries like India.

**References:**

Agiakloglou, C., Drivas, K., & Karamanis, D. (2016). Individual inventors and market potentials: Evidence from US patents. *Science and Public Policy*, *43*(2), 147-156.

Allison, J. E., Sakoda, L. C., Levin, T. R., Tucker, J. P., Tekawa, I. S., Cuff, T., ... & Schwartz, J. S. (2007). Screening for colorectal neoplasms with new fecal occult blood tests: update on performance characteristics. *Journal of the National Cancer Institute*, 99(19), 1462-1470.

Allison, J. R., & Lemley, M. A. (2002). The growing complexity of the United States patent system. *Boston University of Law Review*., 82, 77.

Amram, M. (2005). The challenge of valuing patents and early-state technologies. *Journal of Applied Corporate Finance*, 17(2), 68-81.

Arora, A., Ceccagnoli, M., & Cohen, W. M. (2008). R&D and the patent premium. *International Journal of Industrial Organization*, 26(5), 1153-1179.

Arora, A., Fosfuri, A., & Gambardella, A. (2001). Markets for technology and their implications for corporate strategy. *Industrial and Corporate Change*, 10(2), 419-451.

Azzone, G., & Manzini, R. (2008). Quick and dirty technology assessment: The case of an Italian Research Centre. *Technological Forecasting and Social Change*, 75(8), 1324-1338.

Balasubramanian, N., & Sivadasan, J. (2011). What happens when firms patent? New evidence from U.S. Economic census data. The Review of Economicsand Statistics, 93(1), 126–146.

Baron, J., & Delcamp, H. (2012). The private and social value of patents in discrete and cumulative innovation. *Scientometrics*, *90*(2), 581-606.

Berger, F., Blind, K., & Thumm, N. (2012). Filing behaviour regarding essential patents in industry standards. *Research Policy*, *41*(1), 216-225.

Bessen, J. (2004). *Patent Thickets: Strategic Patenting of Complex Technologies*. Research on Innovation.

Bessen, J. (2004). *Patent thickets: Strategic patenting of complex technologies* (No. 0401).

Bessen, J. (2006). Open source software: Free provision of complex public goods. In *The economics of open source software development* (pp. 57-81). Elsevier.

Bessen, J. (2008). The value of US patents by owner and patent social value indicators. *Research Policy*, 37(5), 932-945.

Bessen, J., & Maskin, E. (2009). Sequential innovation, patents, and imitation. *The RAND Journal of Economics,* 40(4), 611-635.

Bloom, N., Schankerman, M., & Van Reenen, J. (2013). Identifying technology spillovers and product market rivalry. *Econometrica*, *81*(4), 1347-1393.

Brant, R. (1990). Assessing proportionality in the proportional odds model for ordinal logistic regression. *Biometrics*, 1171-1178.

Breitzman, A., Thomas, P., & Cheney, M. (2002). Technological powerhouse or diluted competence: Techniques for assessing mergers via patent analysis.R&D Management, 32(1), 1–10.

Carte, N. (2005). The maximum achievable profit method of patent valuation. *International Journal of Innovation and Technology Management*, 2(02), 135-151.

Caviggioli, F., Scellato, G., & Ughetto, E. (2013). International patent disputes: Evidence from oppositions at the European Patent Office. *Research Policy*, *42*(9), 1634-1646.

Chalioti, E., Drivas, K., Kalyvitis, S., & Katsimi, M. (2020). Innovation, patents and trade: A firm-level analysis. Canadian Journal of Economics/Revuecanadienne d'économique, 53(3), 1–33.

Chang, Y. H., Lai, K. K., Lin, C. Y., Su, F. P., & Yang, M. C. (2017). A hybrid clustering approach to identify network positions and roles through social network and multivariate analysis. *Scientometrics*, *113*(3), 1733-1755.

Cohen, W. M., Nelson, R. R., & Walsh, J. P. (2000). *Protecting their intellectual assets: Appropriability conditions and why US manufacturing firms patent (or not)* (No. w7552). National Bureau of Economic Research.

Cohen, W. M., Nelson, R., & Walsh, J. P. (2000). Protecting their intellectual assets: Appropriability conditions and why US manufacturing firms patent (or not).

Danish, M. S., Ranjan, P., & Sharma, R. (2020). Valuation of patents in emerging economies: a renewal model-based study of Indian patents. *Technology Analysis & Strategic Management*, *32*(4), 457-473.

Deng, Y. (2007). Private value of European patents. *European Economic Review*, 51(7), 1785-1812.

Drivas, K., & Kaplanis, I. (2020). The role of international collaborations in securing the patent grant. *Journal of Informetrics*, *14*(4), 101093.

Ernst, H. (2001). Patent applications and subsequent changes of performance: evidence from time-series cross-section analyses on the firm level. *Research Policy*, 30(1), 143-157.

Ernst, H., Conley, J., & Omland, N. (2016). How to create commercial value from patents: the role of patent management. *R&D Management*, 46(S2), 677-690.

Gambardella, A., D. Harhoff, and B. Verspagen. 2006. The value of patents. Paper presented at the EPIP Conference, September, Munich, Germany. https://www.fep.up.pt/conferences/earie2005/cd_rom/Session%20V/V.C/Verspagen.pdf

Giuri, P., Mariani, M., Brusoni, S., Crespi, G., Francoz, D., Gambardella, A., & Hoisl, K. (2007). Inventors and invention processes in Europe: Results from the PatVal-EU survey. *Research Policy*, 36(8), 1107-1127.

Grimaldi, M., & Cricelli, L. (2019). Indexes of patent value: a systematic literature review and classification. *Knowledge Management Research & Practice*, 1-20.

Grimaldi, M., Cricelli, L., Di Giovanni, M., & Rogo, F. (2015). The patent portfolio value analysis: A new framework to leverage patent information for strategic technology planning. *Technological Forecasting and Social Change*, *94*, 286-302.

Hall, B. H. and Ziedonis, R. H. (2007). 'An empirical analysis of patent litigation in the semiconductor industry'. Working Paper, *National Bureau of Economic Research*. https://eml.berkeley.edu//~bhhall/papers/HallZiedonis07_PatentLitigation_AEA.pdf

Hall, B. H., Jaffe, A. B., & Trajtenberg, M. (2000). Market value and patent citations: A first look (Working Paper No.7741). *National Bureau of Economic Research*. https://www.nber.org/papers/w7741

Hall, B. H., Jaffe, A. B., & Trajtenberg, M. (2001). The NBER patent citation data file: Lessons, insights and methodological tools (Working Paper No. 8498). *National Bureau of Economic Research*. https://www.nber.org/papers/w8498

Hall, B. H., Jaffe, A., & Trajtenberg, M. (2005). Market value and patent citations. *RAND Journal of Economics,* 16-38.

Hall, B. H., Thoma, G., & Torrisi, S. (2007). The market value of patents and R&D: evidence from European firms. In *Academy of Management Proceedings* 1(1-6).

Hall, B. K. (2003). Evo-Devo: evolutionary developmental mechanisms. *International Journal of Developmental Biology*, 47(7-8), 491-495.

Harhoff, D., Narin, F., Scherer, F. M., & Vopel, K. (1999). Citation frequency and the value of patented inventions. *Review of Economics and statistics*, *81*(3), 511-515.

Harhoff, D., Scherer, F. M., & Vopel, K. (2003). Citations, family size, opposition and the value of patent rights. *Research Policy*, 32(8), 1343-1363.

Hausman, J., & Leonard, G. K. (2006). Real options and patent damages: the legal treatment of non-infringing alternatives, and incentives to innovate. *Journal of Economic Surveys*, 20(4), 493-512.

Hikkerova, L., Kammoun, N., & Lantz, J. S. (2014). Patent life cycle: new evidence. *Technological Forecasting and Social Change*, *88*, 313-324.

Hsieh, C. H. (2013). Patent value assessment and commercialization strategy. *Technological Forecasting and Social Change*, *80*(2), 307-319.

Jee, S. J., Kwon, M., Ha, J. M., & Sohn, S. Y. (2019). Exploring the forward citation patterns of patents based on the evolution of technology fields. *Journal of Informetrics*, *13*(4), 100985.

Jensen, P. H., Thomson, R., & Yong, J. (2011). Estimating the patent premium: evidence from the Australian inventor survey. *Strategic Management Journal*, 32(10), 1128-1138.

Kabore, F. P., & Park, W. G. (2019). Can patent family size and composition signal patent value?. *Applied Economics*, *51*(60), 6476-6496.

Kingston, W. (2001). Innovation needs patents reform. *Research Policy*, *30*(3), 403-423.

Klaila, D., & Hall, L. (2000). Using intellectual assets as a success strategy. *Journal of Intellectual Capital*, *1*(1), 47-53.

Kulatilaka, N., & Marcus, A. J. (1992). Project valuation under uncertainty: when does DCF fail?. *Journal of Applied Corporate Finance*, 5(3), 92-100.

Kumar, V., Lai, K. K., Chang, Y. H., Bhatt, P. C., & Su, F. P. (2020). A structural analysis approach to identify technology innovation and evolution path: a case of m-payment technology ecosystem. *Journal of Knowledge Management*.

Kuznets, S. (1962). Inventive activity: Problems of definition and measurement. In The rate and direction of inventive activity: Economic and Social Factors (pp. 19-52). *Princeton University Press.*

Lai, K. K., Bhatt, P. C., Kumar, V., Chen, H. C., Chang, Y. H., & Su, F. P. (2021). Identifying the impact of patent family on the patent trajectory: A case of thin film solar cells technological trajectories. *Journal of Informetrics*, *15*(2), 101143.

Lai, K.-K., Chen, H.-C., Chang, Y.-H., Kumar, V., & Bhatt, P. C. (2020). A structured MPA approach to explore technological core competence, knowledgeflow, and technology development through social network patentometrics. Journal of Knowledge Management.

Lanjouw, J. O., & Schankerman, M. (2001). Characteristics of patent litigation: a window on competition. *RAND Journal of Economics*, 129-151.

Lanjouw, J. O., & Schankerman, M. (2004). Patent quality and research productivity: Measuring innovation with multiple indicators. *The Economic Journal*, *114*(495), 441-465.

Lanjouw, J. O., & Schankerman, M. (2004). Patent quality and research productivity: Measuring innovation with multiple indicators. *The Economic Journal*, 114(495), 441-465.

Lanjouw, J. O., Pakes, A., & Putnam, J. (1998). How to count patents and value intellectual property: The uses of patent renewal and application data. *The Journal of Industrial Economics*, 46(4), 405-432.

Lanjouw, J., & Schankerman, M. (1999). The quality of ideas: measuring innovation with multiple indicators.
26

Levin, R. C., Klevorick, A. K., Nelson, R. R., Winter, S. G., Gilbert, R., & Griliches, Z. (1987). Appropriating the returns from industrial research and development. *Brookings papers on economic activity*, *1987*(3), 783-831.

Levin, R. C., Klevorick, A. K., Nelson, R. R., Winter, S. G., Gilbert, R., & Griliches, Z. (1987). Appropriating the returns from industrial research and development. *Brookings papers on economic activity*, *1987*(3), 783-831.

Machado, M. P. (2004). A consistent estimator for the binomial distribution in the presence of "incidental parameters": an application to patent data. *Journal of Econometrics*, 119(1), 73-98.

Mann, R. J., & Sager, T. W. (2007). Patents, venture capital, and software start-ups. Research Policy, 36(2), 193–208.

Mansfield, E. (1986). Patents and innovation: an empirical study. *Management science*, *32*(2), 173-181.

Marco, A. C. (2005). The option value of patent litigation: Theory and evidence. *Review of Financial Economics*, 14(3-4), 323-351.

McCullagh, P. (1980). Regression models for ordinal data. *Journal of the Royal Statistical Society: Series B* (Methodological), 42(2), 109-127.

Meng, R. (2008). A patent race in a real options setting: Investment strategy, valuation, CAPM beta, and return volatility. *Journal of Economic Dynamics and Control*, 32(10), 3192-3217.

Moore, K. A. (2005). Worthless Patents. *Berkeley Technology Law Journal*, 20, 1521.

Nagaoka, S. (2005). *Patent quality, cumulative innovation and market value: Evidence from Japanese firm level panel data* (No. 05-06). Institute of Innovation Research, Hitotsubashi University.

Neuhäusler, P., & Frietsch, R. (2013). Patent families as macro level patent value indicators: applying weights to account for market differences. *Scientometrics*, *96*(1), 27-49.

Oriani, R., & Sobrero, M. (2008). Uncertainty and the market valuation of R&D within a real options logic. *Strategic Management Journal*, 29(4), 343-361.

Orsenigo, L., & Sterzi, V. (2010). Comparative study of the use of patents in different industries. *Knowledge, Internationalization and Technology Studies (KITeS)*, *33*.

Pakes, A., & Schankerman, M. (1984). The rate of obsolescence of patents, research gestation lags, and the private rate of return to research resources. In R&D, patents, and productivity (pp. 73-88). *University of Chicago Press.*

Park, G., & Park, Y. (2006). On the measurement of patent stock as knowledge indicators. *Technological Forecasting and Social Change*, 73(7), 793-812.

Reitzig, M. (2004). Improving patent valuations for management purposes—validating new indicators by analyzing application rationales. *Research Policy*, 33(6-7), 939-957.

Rycroft, R. W., & Kash, D. E. (1999). *The complexity challenge: Technological innovation for the 21st century*. Burns & Oates.

Schankerman, M., Pakes, A., 1986. Estimates of the value of patent rights in European countries during the post-1950 period. *The Economic Journal* 96, 1052–1076.

Scherer, F. M., Herzstein, S. E., Dreyfoos, A., Whitney, W., Bachmann, O., Pesek, P., ... & Galvin, J. J. (1959). Patents and the Corporation: A Report on Industrial Technology under Changing Public Policy: Privately Published.

Schettino, F., Sterlacchini, A., & Venturini, F. (2013). Inventive productivity and patent quality: Evidence from Italian inventors. *Journal of policy modeling*, *35*(6), 1043-1056.

Schmoch, U. (2008). Concept of a technology classification for country comparisons. Final report to the world intellectual property organisation (wipo), WIPO. https://www.wipo.int/export/sites/www/ipstats/en/statistics/patents/pdf/wipo_ipc_technology.pdf

Serrano, C. J. (2011). Estimating the gains from trade in the market for innovation: Evidence from the transfer of patents(Working Paper No.17304). *National Bureau of Economic Research*. https://www.nber.org/papers/w17304

Shapiro, C. (2000). Navigating the patent thicket: Cross licenses, patent pools, and standard setting. *Innovation policy and the economy*, *1*, 119-150.

Sherry, E. F., & Teece, D. J. (2004). Royalties, evolving patent rights, and the value of innovation. *Research Policy*, 33(2), 179-191.

Singh, J., & Fleming, L. (2010). Lone inventors as sources of breakthroughs: Myth or reality? Management Science, 56(1), 41–56.

Squicciarini, M., Dernis, H., & Criscuolo, C. (2013). Measuring patent quality: indicators of technological and economic value. Organisation for Economic Co-operation and Development (OECD). https://dx.doi.org/10.1787/5k4522wkw1r8-en

Teece, D. J. (1986). Profiting from technological innovation: Implications for integration, collaboration, licensing and public policy. *Research policy*, *15*(6), 285-305.

Thoma, G. (2014). Composite value index of patent indicators: Factor analysis combining bibliographic and survey datasets. *World Patent Information*, *38*, 19-26.

Thomas, P. (1999). The effect of technological impact upon patent renewal decisions. *Technology Analysis & Strategic Management*, *11*(2), 181-197.

Tong, X., & Frame, J. D. (1994). Measuring national technological performance with patent claims data. *Research Policy*, 23(2), 133-141.

Trappey, A. J., Trappey, C. V., Wu, C. Y., & Lin, C. W. (2012). A patent quality analysis for innovative technology and product development. *Advanced Engineering Informatics*, *26*(1), 26-34.

Tsai, J. M., Chang, C. C., & Hung, S. W. (2018). Technology acquisition models for fast followers in high-technological markets: an empirical analysis of the LED industry. *Technology Analysis & Strategic Management*, *30*(2), 198-210.

Van Zeebroeck, N. (2011). The puzzle of patent value indicators. *Economics of innovation and new technology*, *20*(1), 33-62.

Vassolo, R. S., Anand, J., & Folta, T. B. (2004). Non-additivity in portfolios of exploration activities: A real options-based analysis of equity alliances in biotechnology. *Strategic Management Journal*, 25(11), 1045-1061.

Von Wartburg, I., & Teichert, T. (2008). Valuing patents and licenses from a business strategy perspective–Extending valuation considerations using the case of nanotechnology. *World Patent Information*, 30(2), 106-114.

Williams, R. (2006). Generalized ordered logit/partial proportional odds models for ordinal dependent variables. *The Stata Journal*, 6(1), 58-82.